\documentclass[aps,prl,twocolumn,amsmath,amssymb,amsfonts]{revtex4-2}
\usepackage{epsfig}
\usepackage{graphicx}
\usepackage{dcolumn}
\usepackage{bm}
\newcommand{\be}{\begin{equation}}
\newcommand{\ee}{\end{equation}}
\newcommand{\bea}{\begin{eqnarray}}
\newcommand{\eea}{\end{eqnarray}}
\usepackage{color}

\begin{document}

\title{A direct method of continuous unwrapping the phase from an interferogram image}
\author{V. Berejnov$^1$} \email{berejnov@gmail.com} 
\author{B.Y. Rubinstein$^2$}
\affiliation{$^1$ 6575 St. Charles Pl, Burnaby, V5H3W1, Canada}
\affiliation{$^2$ Stowers Institute for Medical Research 1000 50th St., Kansas City, MO 64110, U.S.A.}
\date{\today}

\begin{abstract}
A new method recovers phase difference of interfering wavefronts 
from a pattern of interference fringes, avoiding discontinuity problem. 
The continuous phase is a solution of the first order differential 
equation of the interferogram function computed from the fringe intensity 
profile selected along the pathway over the interferogram.
\end{abstract}


\maketitle




Unwrapping the phase from an interferogram is a common process needed for interpretion of the object wavefront shape 
\cite{Georges2019}. 
In general, the interferogram could be treated as a function with a phase as an argument. Obtaining the phase from 
the interferogram function requires solving a trigonometric equation by inversion \cite{Georges2019}, 
outputting the phase in the so-called wrapped form. 
The wrapped phase is periodic function exhibiting a $2\pi$ type of discontinuity and 
spatially aligned with the interference fringes on the interferogram. 
The process of unwrapping, {\it i.e.}, making the countinuous phase from 
the $2\pi$ pieces is challenging due to this discontinuity, because it requires to track where on the interferogram 
the pieces of the phase must be shifted by $2\pi$. The current state of solving the  $2\pi$ discontinuity problem is based on 
the automatic algorithmic approach and require to recognising, identifying and indexing the phase pieces over the interferogram and 
then manipulating them to remove each individual discontinuity, thereby obtaining the absolute phase function. 
A number of such methods for piecewise phase unwrapping (PPU) are known and could be found 
elsewhere \cite{Judge1994,Reid1986,Malacara2005}.

However, it is possible to find a different mathematical treatment, bypass the trigonomentric approach, solve the phase unwrapping task 
and obtain the continuous phase directly from the interferogram. This formulation eliminates the $2\pi$ discontinuity 
problem, giving access to operate with the phase as a whole piece in its absolute and continuous form. 
Below we present the new method of continuous phase unwrapping (CPU) and give an example comparing both approaches.

{\it Definitions}. An interferogram is defined as an image containing a pattern of bright (constructive) and dark (destructive)
fringes 
projected from the interfering space to the image plane; see an example in Fig.1a. The fringe pattern is 
characterized by the grey values $G(\bm r)$ of the image pixels $\bm r = (x,y)$ oscillating between 
minimal (black) and maximal (white) pixel intensities over the chosen path of the interferogram image, Fig.1b. The function $G(\bm r)$ 
contains the information of the phase difference $\theta(\bm r)$ of the interfering wavefronts of the beams. 
When a wavefront of the first beam is fixed (used as a reference), the whole phase difference is attributed to 
the wavefront of the second beam (called the object beam). The object beam, in this case, is treated as a 
probe to measure or characterize the properties of physical objects: topography of reflecting surfaces, 
shapes of lenses, thickness of thin transparent films, and so on.

The phase difference $\theta$, the optical path difference (OPD) $\delta$ of the object wavefront with respect to the referenced one, 
and the wavelength $\lambda$ satisfy a relationship
\begin{equation}
\theta = 2\pi\delta/\lambda.
\label{eq1}
\end{equation}
If the phase difference $\theta(\bm r)$ is recovered from the grey function $G(\bm r)$, then Eq.(\ref{eq1}) delivers 
the OPD function $\delta(\bm r)$ for given  $\lambda$. OPD characterises the above-mentioned physical properties 
of the interference object once the refraction coefficients of the media where the wavefronts propagate are known.

Following \cite{Judge1994} assume a general form of the function $G$ suitable for the interference experiment 
with a single pass of the object beam producing the interference fringes of infinite width
\begin{equation}
G = A + B \cos\theta,
\label{eq2}
\end{equation}
where the grey function $G$, the coefficients $A,B$, and the phase $\theta$ are the functions of a particular point $\bm r$ on the interferogram. While $G$  oscillates in space, the coefficients $A,B$ are slowly varying functions of spatial coordinates \cite{Judge1994}. Depending on the experimental conditions the coefficients $A,B$ represent a background illumination and an amplitude of the recorded light modulation, respectively. 
For simplicity we consider one-dimensional (1D) functions $\theta(x)$ and $G(x)$.

{\it PPU method example}. The PPU approach for unwrapping $\theta(x)$ from $G(x)$ provides a discontinuous 
form of $\theta(x)$ because $\theta$ being an argument of the cosine function is obtained directly from Eq.(\ref{eq2}) \cite{Okada2007}
\begin{equation}
\theta = 
\arccos \frac{G-A}{B}\;.
\label{eq3}
\end{equation}
Eq.(\ref{eq3}) delivers the piecewise form of the phase $\theta$, where each phase piece $\theta_i$ is 
defined within the interval $0 \le \theta_i \le \pi$ corresponding to two adjacent white and black 
fringes (and {\it vice versa}); see an example in Fig.1c. Eq.(\ref{eq3}) results in the oscillating or 
wrapped form of $\theta$ which must be unwrapped, providing a continuous form of $\theta$ which has a 
physically meaningful shape. The process of unwrapping for the example in Fig.1c requires identification of
the types of the pieces monotonous function $\theta_i(x)$ \cite{Okada2007}. Then, select one type (say with odd 
indices, $\theta_1$, $\theta_3$ in Fig.1c) as “true” and convert the opposite type pieces 
(with even indices $\theta_2$, $\theta_4$ in Fig.1c) into the “true” type by reflecting them as shown in Fig.1c. 
Then shift the converted pieces with respect the preceding ones and obtain the joined 
pieces $\theta_{12}$ and $\theta_{34}$ corresponding now to $2\pi$ discontinuity interval 
and representing only the “true” type of phase change. The continuous shape of the $\theta$ 
profile would be the result of sequentially shifting 
all “true” pairs $\theta_{i,i+1}$ by $2\pi$ with respect to the preceding “true” pairs. All the 
above requires application of the appropriate pattern recognition and analysis methods to the 
wrapped $\theta$ phase, allowing recognition and indexing the fringes, splitting into pieces, 
converting, shifting, and then obtaining the continuous form of the unwrapped $\theta$ 
phase by manipulating the adjacent pairs. The current PPU methods are much more 
sophisticated than the above, but they are all designed to remove $2\pi$ discontinuities in a piecewise manner.

\begin{figure}[h!]
\begin{center}
\psfig{figure=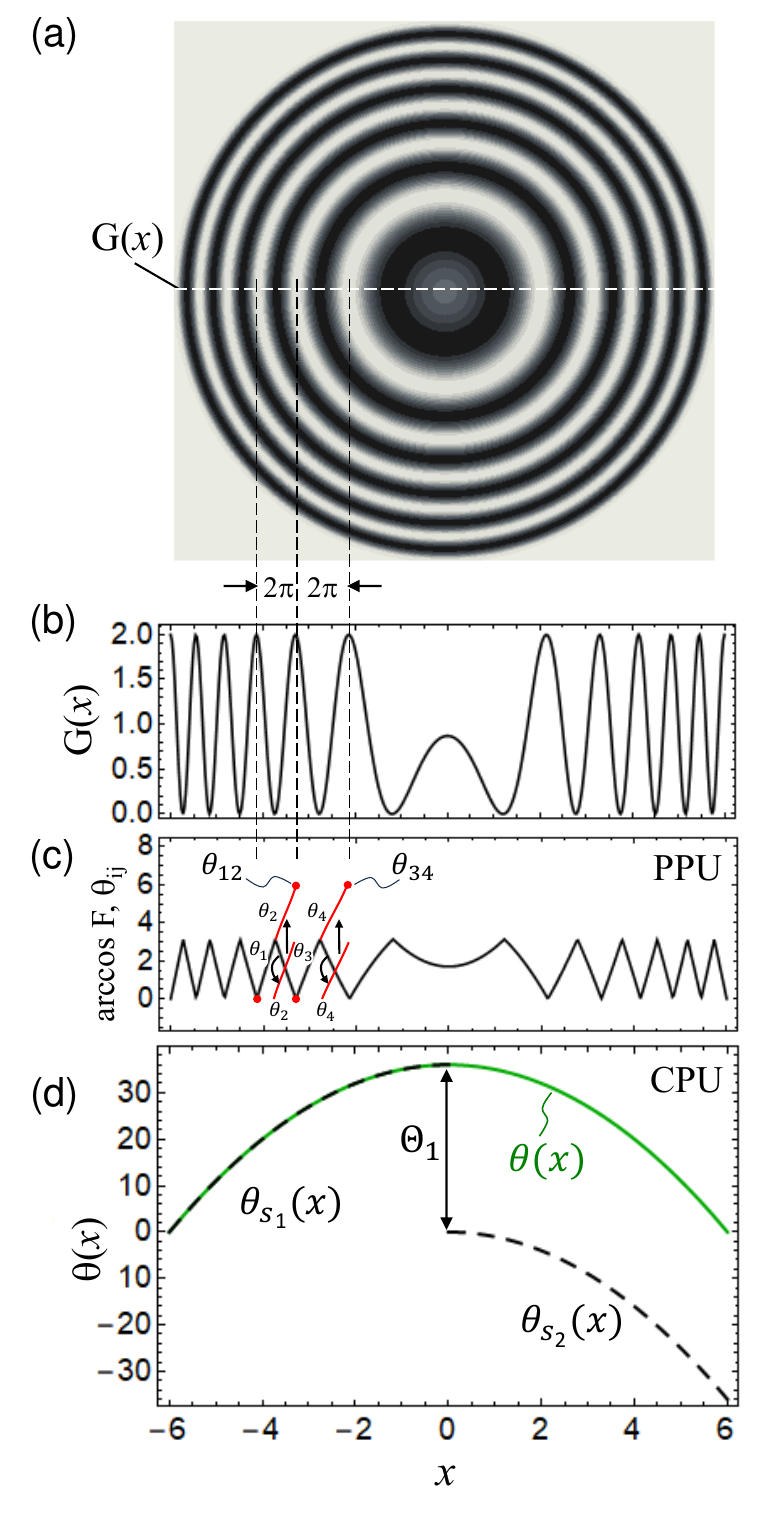,width=8.0cm}
\end{center}
\caption{The interferogram (a) with the grey function 
$G(x,y)=1+\cos(6^2-x^2-y^2)$. (b) is a profile of the function 
$G(x,0)=G(x)=1+\cos(6^2-x^2)$ sampled along the straight path $y=0$ for analysis. 
(c) illustrates the PPU method: a profile of $\arccos (G-A)/B$ in Eq.(\ref{eq3}) is 
represented by the black curve, 
the phase pieces $\theta_{i}$ correspond to $\pi$ intervals and the pieces joined in 
pairs $\theta_{i,i+1}$ to $2\pi$ intervals, respectively. 
The black arrows show a method of converting the even pieces of 
phase $\theta_{2}$, $\theta_{4}$ into the ``true'' directed pieces and 
then shifting them with respect to the ``true'' preceeding pieces $\theta_{1}$ 
and $\theta_{3}$ to obtain the ``true'' pair-joined pieces $\theta_{12}$, $\theta_{34}$, see the text. 
(d) illustrates the CPU method: the black dashed curves present the phases 
for intervals $s_1$ and $s_2$; the functions $\theta_{s_1}(x)$, $\theta_{s_2}(x)$ 
and the phase shift $\Theta_{1}$ all are computed from Eq.(\ref{eq7}). 
The green curve represents the resulting phase $\theta(x)$. 
All graphics are aligned along $x$ coordinate.}
\label{fig1}
\end{figure}

{\it CPU method}. Consider an arbitrary 2D interferogram characterized by a digital grey 
function $G(x)$ known for each pixel of the entire area of the interferogram. Select a 
straight pathway parallel to the $x$-axis to obtain a 1D grey function of interest $G(x)$ 
defined over the interval $x_{min} \le x \le x_{max}$. Introduce an interferogram 
function $F$ -- a relationship between the phase difference $\theta(x)$ and the grey 
function $G(x)$ obtained from a particular experimental setup. Rewrite Eq.(\ref{eq2}) 
to obtain $F$ corresponding to a single passage of the object beam through the 
interfering medium with a unit refraction index
\begin{equation}
\cos \theta = F(x) = 
\frac{G(x)-A}{B}\;.
\label{eq4}
\end{equation}
Differentiating Eq.(\ref{eq4}) with respect to $x$ we find
\begin{equation}
- \theta'(x)  \sin \theta = F'(x).
\label{eq5}
\end{equation}
Use Eqs.(\ref{eq4},\ref{eq5}) in Pythagorean identity $\sin^2\alpha+cos^2\alpha=1$ 
to eliminate the trigonometric functions from consideration and obtain
\begin{equation}
\theta'(x)^2  = \frac{F'(x)^2}{1-F^2(x)}
\label{eq6}
\end{equation}
Note that in the differential equation Eq.(\ref{eq6})
the phase difference is already unwrapped, {\it i.e.}, $\theta(x)$ is defined over the entire interval $x_{min} \le x \le x_{max}$. 
Applying the relations $\sqrt{f^2}=|f|$, where $|.|$ denotes a modulus, and $|f(x)|=\mbox{sgn}(f) f \le 0$, 
where $\mbox{sgn}(.)$ stands for the sign function returning $\pm1$, 
write the solution of Eq.(\ref{eq6}) as
\begin{equation}
\theta(x)  = \mbox{sgn}(\theta'(x)) \int_{x_{min}}^x \frac{|F'(\xi)|}{\sqrt{1-F^2(\xi)}}\; d\xi.
\label{eq7}
\end{equation}
The function $F(x)$ and its derivative can be obtained from Eq.(\ref{eq4}) by 
employing the interference pattern $G(x)$. The integrand in Eq.(\ref{eq7})
at some points in the integration interval represents the 0/0 indeterminate, 
however the computation shows that terms contributing to a (possible) divergence cancel out. 
The function $\mbox{sgn}(\theta'(x))$ could be obtained once the extremum points $x_i$ 
of $\theta(x)$ are found from the equation $\theta'(x)=0$. As at these points both 
sides of Eq.(\ref{eq6}) vanish, the equation for the extrema $x_i$ reads
\begin{equation}
\frac{F'(x)^2}{1-F^2(x)}=0.
\label{eq8}
\end{equation}
In general case Eq.(\ref{eq8}) can be solved numerically. For a sequence of $n$ 
roots $x_i,\ 1\le i \le n$, of Eq.(\ref{eq8}) the whole interval should be divided into a sequence 
of $n+1$ segments $s_i:\{x_{i-1}\le x \le x_{i}\},\ 1\le i \le n+1$ 
where each segment is characterized by a specific value of 
$\sigma_i=\mbox{sgn}(\theta'(x))$ with $\ x \in s_i$. The sign alternates between adjacent segments. 
Thus, the whole sign sequence is determined by the $\sigma_1$ in the first 
segment $s_1: \{x_0 =  x_{min} \le x \le x_1\}$ and it does not affect the shape of the 
unwrapped phase. The integral of Eq.(\ref{eq7}) must be applied for each individual 
segment $s_i$ taking into account the phase of the preceding segment.

{\it CPU method example}. To illustrate the CPU method consider the phase difference 
defined as a one dimensional even parabolic function
$\theta(x)=R^2-x^2$, in the interval $-R \le x \le R$ with constant $R$. 
Use Eq.(\ref{eq2}) to obtain the grey function 
\begin{equation}
G = A + B \cos(R^2-x^2),
\label{eq9}
\end{equation}
where $A,B$ are constants. The goal is to evaluate Eq.(\ref{eq7}) and find the phase $\theta(x)$. 
For the interferogram function $F$ is given by
\begin{equation}
F(x) = 
(G(x)-A)/B = \cos(R^2-x^2)\;,
\label{eq10}
\end{equation}
leading to $F'(x) = 2x \sin(R^2-x^2)$.
Eq.(\ref{eq8}) gives an equation for the roots $4x^2=0$. 
In the interval $-R \le x \le R$ there is a single root $x_1=0$, being an extremum point of the 
phase function, so it produces only two sign segments. In the first segment $s_1:\{-R \le x \le 0\}$ the 
sign $\sigma_1=+1$, the integrand of Eq.(\ref{eq7}) is $-2x$, and 
the phase $\theta_1(x)$ for this interval reads
\begin{equation}
\theta_1(x) = \theta_{s_1}(x) = 
-2 \int_{-R}^x \xi d\xi = R^2-x^2,
\label{eq11}
\end{equation}
where $\theta_{s_1}(x)$ denotes the phase per segment.
The phase shift for the whole first segment $\Theta_1=R^2$. For the second 
segment $s_2:\{0 \le x \le R\}$ we have $\sigma_2=-1$, the integrand is $2x$, 
and the phase for this interval $\theta_2(x)$ evaluates to
\begin{equation}
\theta_2(x) = 
\Theta_1 + \theta_{s_2}(x) = R^2 -2 \int_{0}^x \xi d\xi = R^2-x^2,
\label{eq12}
\end{equation}
Uniting the intervals we obtain the final phase $\theta(x) = \theta_1(x) \cup \theta_2(x) = R^2-x^2$
in the entire interval $-R \le x \le R$ reproducing the identity of 
the input and recovered phase, see Fig.1d.

The proposed method can be applied for $G(x)$ defined numerically (as an array), an 
unknown phase $\theta(x)$ may include more than one extreme point. In this case, the roots of 
Eq.(\ref{eq8}) and the integral Eq.(\ref{eq7}) are computed numerically, once the process of 
digitizing the experimental interference pattern provides a sufficient number of points per the 
narrowest fringe in the $G(x)$ function used as an input for the interferogram function $F$. 
Note, this method allows us to obtain the phase $\theta(x)$ profile from the experimental $G(x)$ 
profile with the accuracy of the first segment $s_1$. The sign $\sigma_1$ for the segment $s_1$ is set 
arbitrarily, then one obtains two $\theta(x)$ profiles of the same shape but mirror 
reflected to each other with respect to the $x$-axis. One might
choose the correct shape by using the experimental insights usually existing for some 
characteristic points on the interferogram.

The presented method of continuous phase unwrapping (CPU) could be considered as 
complimentary to the existing one – piecewise phase unwrapping (PPU). Regarding the 
fundamental difference between the methods, the CPU method uses the numerical calculus approach
providing access to the whole phase function without $2\pi$ discontinuities, while PPU applies automatic
algorithmic manipulations to the phase pieces to remove these $2\pi$ discontinuities. 
The CPU approach may offer a new perspective in developing 
the methods for interferogram analysis. For example, the CPU method allows one to obtain
the phase and OPD profiles \cite{BerejnovLi2010} without the extensive programming required for the PPU method.


\end{document}